\documentclass[fleqn,usenatbib]{mnras}
\usepackage{longtable}
\usepackage{amssymb}
\usepackage{amsmath}
\usepackage[T1]{fontenc}

\DeclareRobustCommand{\VAN}[3]{#2}
\let\VANthebibliography\thebibliography
\def\thebibliography{\DeclareRobustCommand{\VAN}[3]{##3}\VANthebibliography}

\usepackage{graphicx}	% Including figure files
\usepackage{amsmath}	% Advanced maths commands
\usepackage{amssymb}	% Extra maths symbols

\newcommand{\Mjup}{\mbox{M$_{\mathrm{jup}}$}}
\newcommand{\Msun}{\mbox{M$_{\odot}$}}
\newcommand{\Rsun}{\mbox{R$_{\odot}$}}

\newcommand{\Mwd}{\mbox{$M_{\mathrm{WD}}$}}
\newcommand{\Mbd}{\mbox{$M_{\mathrm{BD}}$}}

\newcommand{\Ebin}{\mbox{$E_{\mathrm{bin}}$}}
\newcommand{\Eorb}{\mbox{$E_{\mathrm{orb}}$}}

\newcommand{\gppr}{\stackrel{>}{\scriptstyle \sim}}
\newcommand{\gappr}{\raisebox{-0.4ex}{$\gppr$}}

\title[Close WD+BDs: evidence for a low CE efficiency]{Close detached white dwarf + brown dwarf binaries: further evidence for low values of the common envelope efficiency}

\author[M. Zorotovic \& M.R. Schreiber]{
Monica Zorotovic,$^{1}$\thanks{E-mail: monica.zorotovic@uv.cl (MZ)},
Matthias.R. Schreiber,$^{2,3}$
\\
$^{1}$Instituto de F\'isica y Astronom\'ia, Universidad de Valpara\'iso, Av. Gran Breta\~na 1111, Valpara\'iso, Chile\\
$^{2}$Departamento de F{\'i}sica, Universidad T\'ecnica Federico Santa Mar\'ia, Av. España 1680, Valpara{\'i}so, Chile.\\
$^{3}$Millennium Nucleus for Planet Formation, NPF, Valpara{\'i}so, Chile\\
}

\date{Accepted 2022 April 20. Received 2022 April 11; in original form 2022 March 14}

\pubyear{2022}

\begin{document}
\label{firstpage}
\pagerange{\pageref{firstpage}--\pageref{lastpage}}
\maketitle

\begin{abstract}
Common envelope evolution is a fundamental ingredient in our understanding of the formation of close binary stars containing compact objects which includes the progenitors of type Ia supernovae, short gamma ray bursts and most stellar gravitational wave sources. To predict the outcome of common envelope evolution we still rely to a large degree on a simplified energy conservation equation. Unfortunately, this equation contains a theoretically rather poorly constrained efficiency parameter ($\alpha_{\mathrm{CE}}$) and, even worse, it is unclear if energy sources in addition to orbital energy (such as recombination energy) contribute to the envelope ejection process. In previous works we reconstructed the evolution of observed populations of post common envelope binaries (PCEBs) consisting of white dwarfs with main sequence star companions and found indications that the efficiency is rather small ($\alpha_{\mathrm{CE}}\simeq0.2-0.3$) and that extra energy sources are only required in very few cases. Here we used the same reconstruction tool to investigate the evolutionary history of a sample of observed PCEBs with brown dwarf companions. In contrast to previous works, we found that the evolution of observationally well characterized PCEBs with brown dwarf companions can be understood assuming a low common envelope efficiency ($\alpha_{\mathrm{CE}}=0.24-0.41$), similar to that required to understand PCEBs with main sequence star companions, and that contributions from recombination energy are not required. We conclude that the vast majority of PCEBs form from common envelope evolution that can be parameterized with a small efficiency and without taking into account additional energy sources. 
\end{abstract}


%%%%%%%%%%%%%%%%%%%%%%%%%%%%%%%%%%%%%%%%%%%%%%%%%%
%%%%%%%%%%%%%%%%% INTRO %%%%%%%%%%%%%%%%%%
%%%%%%%%%%%%%%%%%%%%%%%%%%%%%%%%%%%%%%%%%%%%%%%%%%

\section{Introduction}

Binary systems containing a white dwarf (WD) orbited by a close companion are expected to form after the ejection of the WD's progenitor envelope during the giant phase in a common-envelope (CE) event \citep[e.g.,][]{Paczynski76,webbink84-1,Ivanova13}. The variety of systems descending from this evolutionary path includes fascinating objects like cataclysmic variables \citep{warner95}, low-mass X-ray binaries \citep{paterson84}, super soft X-ray sources \citep{yungelson96} or double degenerate WD binaries \citep{webbink84-1}. The last two are considered possible supernova Ia progenitors through the single \citep{han04} and double \citep{iben+tutukov84-1} degenerate channels, respectively. Despite its relevance, CE evolution is a very poorly understood phase, usually approximated by simple equations relating the total energy \citep[e.g.,][]{iben+livio93,yungelson94,PRH03} or angular momentum \citep{nelemans+tout05} before and after the CE phase. 

The most widely used approach, both for the simulation of detached post-common envelope binaries (PCEBs) and for the reconstruction of the evolutionary history of the observed PCEB sample, is the \textit{energy formalism} developed by \citet{webbink84-1}. In this approximation, the binding energy of the envelope (\Ebin) is compared to the change in orbital energy due to the CE phase ($\Delta$\Eorb) scaled with an efficiency $\alpha_{\mathrm{CE}}$ which corresponds to the fraction of the change in orbital energy that is used to unbind the envelope, i.e.
\begin{equation}
\Ebin = \alpha_{\mathrm{CE}} \Delta\Eorb.
\end{equation}
\label{eq1}
Several attempts have been made to constrain the CE efficiency by theoretically  reconstructing the evolutionary history of observed PCEBs as well as by comparing the predictions of simulations with observational samples. A small CE efficiency ($\alpha_{\mathrm{CE}}\sim0.2-0.3$) has been derived from observations of WDs containing a close detached low-mass (spectral type M) main-sequence companion \citep[WD+dM, e.g.,][]{zorotovicetal10-1,Toonen13,Camacho14}. This same small efficiency allows to understand the formation of most detached PCEBs with more massive main-sequence companions of spectral type A, F, G or K \citep[WD+AFGK systems,][]{parsons2015,Hernandez21,hernandezetal22}. The only exceptions are IK\,Peg \citep{landsman93,zorotovicetal10-1} and KOI\,3278 \citep{kruseagol14,zorotovicetal14-2}. Both have periods longer than most PCEBs (21.7 and 88.18 days, respectively) and require an unrealistic high CE efficiency (larger than 1) or the inclusion of a small contribution from extra energy sources, such as hydrogen recombination energy, in order to explain their current orbital configuration. 

To progress with our understanding of CE evolution it is important to investigate the evolutionary history of potential PCEBs over a wide range of secondary star masses as this may either allow to derive constraints on a universal value for the CE efficiency or indicate the existence of processes that are not included in the simple energy equation. Previous studies have suggested that detached PCEBs composed of a WD and a brown dwarf companion (WD+BDs) might need a larger CE efficiency ($\alpha_{\mathrm{CE}}>1$) in order to reconstruct their CE phase \citep[][]{demarco11,davis2012}. However, this suggestion was based on two systems only, and the presence of the companion was later ruled out in one of them, as we will discuss in detail in section \ref{sec:sample}. A decade later, it is time to analyze the current, much larger sample of observed detached WD+BD PCEBs.

Detached WD+BD binaries are hard to detect given the low intrinsic brightness of the brown dwarf (BD), which is easily overshadowed by the brightness of the WD. The first detection of a PCEB containing a BD in close orbit around a WD was presented by \citet{maxtedetal06} and today a dozen such systems have been suggested \citep[e.g.][]{vanroestel21,Kruckow21}. The most useful systems among these binaries are certainly those where the WD is eclipsed by the BD. These systems allow for the precise determination of the orbital and stellar parameters of both stars, which is otherwise almost impossible for the sub-stellar object. Today we know four close eclipsing WD+BD binaries. The first discovery was based on \citet{beuermann13} who strongly suggested that the unseen companion in the eclipsing PCEB CSS21055 is a sub-stellar object. Similar systems were subsequently reported by \citet{parsons17,rappaport17}, \citet{Casewell2020} and very recently \citet{vanroestel21}. 

Although there are still relatively few PCEBs with BD companions known and only for four of them (the eclipsing ones) we have precise parameter measurements, the reconstruction of their evolutionary history can give us a first indication as to whether the engulfment of a BD during the CE phase can be approximated in the same way as for low-mass main-sequence companions. 

In this article we present a list of observationally well characterized close detached WD+BDs and reconstruct their formation history based on the energy formalism for CE evolution in order to derive constraints on the CE efficiency.

\section{Observed sample} 
\label{sec:sample}

\begin{table*}
	\centering
	\caption{Orbital parameters for the WD+BD systems with accurate parameters from the literature. The first four correspond to the eclipsing systems.}	\begin{tabular}{llccccccl} 
    \hline
ID & $P_{\mathrm{orb}}$\,[days] & $\Mwd$\,[\Msun] & $\Mbd$\,[\Msun] & $T_\mathrm{eff,WD}$\,[K] & $t_\mathrm{cool}$\,[Myr] & d[pc] & age [Gyr] & Ref\\
\hline
SDSS\,J1411$+$2009 & 0.0845327526(13)  & 0.53$\pm$0.03     & 0.050$\pm$0.002    & 13\,000$\pm$300 & 260$\pm$20 & 190$\pm$8  & $\gappr$3    & 1,2 \\
SDSS\,J1205$-$0242 & 0.049465250(6)    & 0.39$\pm$0.02     & 0.049$\pm$0.006    & 23\,680$\pm$430 & $\sim50$   & 720$\pm$40 & 2.5 -- 10    & 3 \\
WD\,1032$+$011     & 0.09155899610(45) & 0.4502$\pm$0.0500 & 0.0665$\pm$0.0061  & 9\,950$\pm$150  & 455$\pm$80 & 327$\pm$37 & 5 -- 10      & 4 \\
ZTF\,J0038$+$2030  & 0.4319208(14)     & 0.50$\pm$0.02     & 0.0593$\pm$0.004   & 10\,900$\pm$200 & $\sim400$  & $\sim140$  & $\gappr$8    & 5\\ 
WD\,0137$-$349     & 0.079429939(1)         & 0.39$\pm$0.035    & 0.053$\pm$0.006    & 16,500$\pm$500  & 250$\pm$80 & 102$\pm$3  & $\gappr1$     & 6,7, 8\\
NLTT\,5306         & 0.070750(14)      & 0.44 $\pm$0.04    & >\,0.053$\pm$0.003 & 7\,756$\pm$35   & 710$\pm$50 & 71$\pm$4   & 5 -- 10      & 9\\
SDSS\,J1557$+$0916 & 0.094714708(83)   & 0.447$\pm$0.043   & 0.063$\pm$0.002    & 21\,800$\pm$800 & 33$\pm$5   & 520$\pm$35 & ?            & 10 \\
EPIC\,212235321 & 0.047369569(56)       & 0.47$\pm$0.01     & <\,0.0668          & 24\,490$\pm$194 & 18$\pm$1   &  386.8          & >\,0.7       & 11, 12 \\
\hline
\end{tabular} 
\noindent
\\
References: [1] \citet{beuermann13}; 
[2] \citet{littlefair14}; 
[3] \citet{parsons17}; 
[4] \citet{Casewell2020}; 
[5] \citet{vanroestel21}; 
[6] \citet{maxtedetal06}; 
[7] \citet{burleighetal06}; 
[8] \citet{longstaff17}; 
[9] \citet{steele13};
[10] \citet{Farihietal17};
[11] \citet{casewell18};
[12] \citet{gaia-collab18}
\label{tab:1}
\end{table*}

About a dozen of close detached WD+BDs have been reported in the literature since the first discovery by \citet{maxtedetal06}. In what follows we briefly review the orbital parameters for each of these systems and how they were derived. We divide the sample in eclipsing and non-eclipsing systems, because more accurate parameters can usually be derived for eclipsing systems (especially regarding the BD) and we therefore consider those parameters more reliable. We also clarify why some of the systems suspected to be detached WD+BDs in the literature were left out of the sample we use in this paper. 

\subsection{Eclipsing systems}

The first eclipsing detached WD+BD binary discovered was SDSS\,J141126.20$+$200911.1 (hereafter SDSS\,J1411$+$2009) also known as CSS21055. \citet{drake10} first cataloged it as an eclipsing system and it was later included in the \textit{Sloan Digital Sky Survey (SDSS) DR7 White Dwarf Catalog} \citep{kleinman13} where the WD dominates both spectroscopy and photometry. Based on in-eclipse Bessel I-band observations, \citet{beuermann13} derived an upper limit for the mass of the companion which strongly suggested it was a sub-stellar object. The BD nature was later confirmed by \citet{littlefair14} based on high time resolution light curves of the primary eclipse (in SDSS-$u'g'z'$ filters), near-infrared photometry (in $J$, $H$, and $K_s$ bands) and time-resolved X-Shooter spectroscopy. Based on their high quality data, \citet{littlefair14} were able to constrain the mass and radius of the companion and to refine the parameters for the WD. 

Later, \citet{parsons17} and \citet{rappaport17} reported, almost at the same time, the discovery of the second eclipsing detached WD+BD binary, SDSS\,J120515.80$-$024222.6 (hereafter SDSS\,J1205$-$0242) also known as EPIC\,201283111 or WD\,1202$-$024. The system was discovered in the \textit{K2} project \citep{howelletal14-1} which represents and extension of the Kepler Planet-Detection Mission \citep{Boruckietal2010-1}. The orbital parameters derived in the two discovery papers slightly differ. Throughout this work we use the parameters obtained by \citet{parsons17} because the distance they derived ($720\pm40$\,pc) agrees much better with the parallax ($1.504\pm0.234$\,mas) measured by \textit{Gaia} \citep{brownetal18-1} than the larger distance ($845\pm65$\,pc) obtained by \citet{rappaport17}. 

The third BD eclipsing a WD was identified in the detached system WD\,1032$+$011, also known as SDSS\,J103448.94$+$005201.3. The DA WD was first identified by \citet{vennes02} in the 2dF QSO Redshift Survey \citep{croometal01} and insight of a BD companion was derived by \citet{steele11} who analyzed its magnitudes in SDSS $ugriz'$ filters and UKIDSS $H$ and $K$ bands, finding a near-infrared excess consistent with a companion of spectral type L5. The sub-stellar nature of the companion was recently confirmed by \citet{Casewell2020}, who combined optical light curves with optical and near infrared spectroscopy and derived accurate orbital parameters.

Very recently, \citet{vanroestel21} discovered the fourth eclipsing detached WD with a sub-stellar companion, ZTF\,J003855.0$+$203025.5 (hereafter ZTF\,J0038$+$2030), also known as SDSS\,J003854.98$+$203025.7. The authors searched for eclipsing WDs in the \textit{Gaia} Data Release 2 catalogue from \citet{Gentile2019}. Zwicky Transient Facility \citep[ZTF,][]{Bellm19} light curves of the system show a complete eclipse of short duty cycle in both \textit{g-} and \textit{r-}band and were used by \citet{vanroestel21} to determine the orbital period. They also used archival data from several surveys in order to study the spectral energy distribution, covering a large range in wavelengths from the ultraviolet to far-infrared. The contribution of the companion to the WD is negligible, even in the infrared, which indicates that it is cold and of sub-stellar nature. Follow up high-speed \textit{g-} and \textit{z-}band photometry as well as phase-resolved medium-resolution spectroscopy allowed \citet{vanroestel21} to constrain the masses and radii of both components in the binary system, confirming that the companion to the WD is an old ($\gappr$8\,Gyr) BD. Although the authors refer to it as a \textit{long-period system}, the orbital period is $\sim$10 hours, which is an order of magnitude longer than that of the three previously discovered eclipsing detached WD+BDs but still short enough to be certainly the result of CE evolution. 

The observed orbital parameters for the four eclipsing detached WD+BD binaries currently known are listed in Table\,\ref{tab:1} (first four rows). We note that CE evolution truncates the growth of the core mass for the progenitor of the WD, resulting in a lower WD mass than expected from single star evolution, and therefore the initial-to-final mass relation cannot be used to derive the total age of the systems.
The limits derived for the total ages (last column in Table\,\ref{tab:1}) were normally determined as being consistent with the model mass and radius for the BD. However, the BD in WD1032$+$011 seems to be inflated and therefore \citet{Casewell2020} used another method to estimate the age of the system, based on kinematics.
%based on a comparison of the estimated BDs radii and masses with the predictions of evolutionary models for SDSS\,J1411$+$2009, SDSS\,J1205$-$0242 and ZTF\,J0038$+$2030, while it is based on kinematics in the case of WD1032$+$011, which strongly suggest it belongs to the thick disc of the Galaxy. This is because \citet{Casewell2020} suspect that the BD in WD1032$+$011 is inflated, which makes an age estimation based on the BD radius not reliable.

\subsection{Non-eclipsing systems}

The first evidence of a BD in close orbit around a WD was presented by \citet{maxtedetal06}. The system, WD\,0137$-$349, consists of a low mass WD orbited by a $\sim0.05\,\Msun$ BD with an orbital period of $\sim1.9$\,h. This mass of the BD agrees with the L8 spectral type derived by \citet{burleighetal06} from IR spectroscopy for a cooling age of $\simeq1$\,Gyr. Given that the BD might have been reheated following CE evolution, we consider this age as a lower limit for the age of the system. We note that this system has become a key object for studying irradiated atmospheres of tidally locked and fast rotating atmospheres which led to precise determination of some of the system parameters \citep{casewelletal15,longstaff17,zhouetal22}. 

NLTT\,5306, also known as SDSS\,J013532.98$+$144555.8 or WD\,0132$+$145, was identified simultaneously by \citet{steele11} and \citet{girven11} as a WD with a possible late-type stellar companion or BD, based on infrared excess emission. Follow up spectroscopy of the system allowed \citet{steele13} to confirm the presence of a sub-stellar companion and to derive orbital parameters which are similar to the ones derived for WD\,0137$-$349. The kinematics of the system suggest it belongs to the Galactic thick disc, which implies the system is most likely older than 5\,Gyr. There are also signs of accretion onto the WD from a wind \citep{longstaff19} and the BD seems to be inflated \citep{casewell2020b,buzardetal2022}. 

An intriguing system is SDSS\,J155720.78$+$091624.7 (hereafter SDSS\,J1557$+$0916) which hosts a circumbinary disc as already suggested by \citet{steele11} and later confirmed by \citet{Farihietal17}. 
In addition, the WD is clearly metal polluted and therefore most likely accreting planetary debris from the circumbinary disc. The WD parameters of this system were derived from combining spectral fitting with cooling models of WDs. The mass of the companion was estimated by combining radial velocity measurements of the WD with those derived from emission lines coming from the irradiated hemisphere of the BD.  

EPIC\,212235321 was identified as a WD+BD binary using data from the \textit{K2} mission. The light curve did not show eclipses but a strong reflection effect which permitted \citet{casewell18} to derive rather precise parameters. The strong irradiation of the BD in this very close system also generates metal emission lines coming from the BD which were detected through follow-up spectroscopy. 

\subsection{Systems not included in our reconstruction}
\label{sec:not_incl}

There are other systems in the literature that have been included in samples of possible close detached WD+BDs and that we decided to not consider for our reconstruction, based on their uncertain parameters and/or small probability that they are in fact WD+BD binaries, as including uncertain systems might affect any possible conclusion on CE parameters.

One of such systems is GD\,1400, also known as 	WD\,0145$-$221, which has been suggested to be a WD+BD system (DA+dL6-7) by \citet{farihi+christopher04} and \citet{dobbie05}. \citet{burleigh11} derived an orbital period of $\sim10$\,h, which would place this system within the WD+BD PCEBs with the longest period known, similar to the one derived for ZTF\,J0038$+$2030.
However, the orbital period of this system was never published in a peer reviewed journal and we also did not find recent studies on this object. In \citet{burleigh11} only limits on the masses of the components and no constraints on the cooling age or the total age of the system are provided. 

A close sub-stellar companion around WD\,0837$+$189\footnote{Erroneously named WD\,0837$+$185 in the discovery paper \citep{dobbie2004} and in \citet{casewelletal12}.} (also known as LB\,5959) was suggested by \citet{casewelletal12} within the open star cluster Praesepe. However, the companion is not detected in this system and its existence is only inferred from radial velocity (RV) variations of the WD. Inspecting \citet{casewelletal12} we found that the RV measurements used to determine the orbital period come from combining two subsequent exposures of 20\,minutes which implies that each measurement covered $\sim15$ per cent of the orbit (if the $4.2$\,h orbital period is correct). Unfortunately, no power spectrum is shown and no uncertainty is given for the measured orbital period which makes it somewhat difficult to assess the reliability of the measurement. In addition, it seems that the resulting RV curve does not fit the data extremely well. At the minimum of the RV curve the best fit disagrees with the measurements even taking into account the relatively large uncertainties of each individual RV measurement \citep[][their fig.\,1]{casewelletal12}. Finally, according to \citet{salaris+bedin19} all \textit{Gaia} DR2 parameters for this object, and in particular the estimated errors in magnitudes, do not show any indication of this being a binary system but are consistent with this object being a single star. We therefore decided to not include this system in our sample of certain WD+BD systems. 

SDSS\,J123127.14$+$004132.9 \citep{parsons17}, also known as EPIC\,248368963, was not considered because the companion might not be a BD. The WD mass is rather well known (0.56$\pm$0.07\Msun) and the WD is partially eclipsed by a companion of mass $\leq0.095$\Msun. 

Very recently, \citet{rebassaetal22} presented the hierarchical triple system Gaia\,0007$-$1605, whose inner binary seems to be a WD+BD system with an orbital period of $\sim1$\,day, while the outer component is another WD. 
We decided to not include this system in our reconstruction because its triple nature makes its evolutionary history potentially more complex. 
Given that the close transiting giant planet around WD\,1856$+$534, which also hosts a distant tertiary, has perhaps reached its current orbit due to the Kozai-Lidov effect 
\citep[see e.g.][for a discussion of possible evolutionary pathways]{munoz+petrovic20,oconnor21,lagosetal21-1}, we cannot a priori exclude a similar scenario for Gaia\,0007$-$1605. 

Finally, SDSS\,J121209.31$+$013627.7 \citep{schmidt05} is a magnetic WD with a possibly sub-stellar companion and it was speculated that it might be a detached system \citep{burleighetal06}. However, it seems clear today that it is instead a cataclysmic variable where the WD accretes from the L type companion \citep{stelzer17}. We note that several cataclysmic variables with BD companions are known \citep[e.g.,][]{Hernandez-Santisteban16,Kruckow21} but these systems cannot be used to constrain CE evolution as the mass transfer history and therefore the initial mass of the secondary can neither theoretically nor observationally be determined. 

\vspace{0.5cm}
\noindent Ignoring the systems detailed in Sec.\,\ref{sec:not_incl}, our final sample consists of the eight systems presented in Table\,\ref{tab:1}. In the next section, we reconstruct their evolutionary history with special emphasis on the solutions that are consistent with their minimum ages estimated in the literature. 

%%%%%%%%%%%%%%%%%%%%%%%%%%%%%%%%%%%%%%%%%%%%%%%%%%
%%%% MOVE TABLEs WERE IT LOOKS BETTER IN THE PDF %%
%%%%%%%%%%%%%%%%%%%%%%%%%%%%%%%%%%%%%%%%%%%%%%%%%%

 \begin{table*}
	\centering
	\caption{Range of reconstructed parameters for the close detached WD+BD systems known with accurate parameters, assuming CE evolution without any contribution from recombination energy and requesting $\alpha_\mathrm{CE}$ to not exceed 1. $P_{\mathrm{CE}}$ is the orbital period immediately after the CE phase. $P_{\mathrm{orb,i}}$ and $M_{\mathrm{1,i}}$ are the initial orbital period and initial mass of the primary star (i.e. the progenitor of the WD). $R_\mathrm{1,CE}$ is the radius of the primary star at the onset of the CE phase.}
	\begin{tabular}{llcccc} 
    \hline
    ID & $P_{\mathrm{CE}}$\,[days] & $P_{\mathrm{orb,i}}$\,[days] & $M_{\mathrm{1,i}}$\,[\Msun] & $R_\mathrm{1,CE}$\,[\Rsun] & $\alpha_{\mathrm{CE}}$\\ 
    \hline
\multicolumn{6}{c}{For any total age $\leq$ 10\,Gyr}\\
\hline
SDSS\,J1411$+$2009 & 0.0884218 & 186 - 937 & 1.08 - 2.24 & 112 - 276 & 0.09 - 1 \\
SDSS\,J1205$-$0242 & 0.0509304 & 67 - 200  & 1.07 - 1.86 & 54 - 92   & 0.18 - 1 \\
WD\,1032$+$011     & 0.0983930 & 99 - 605  & 1.08 - 1.86 & 79 - 178  & 0.06 - 1 \\
ZTF\,J0038$+$2030  & 0.4323880 & 591 - 925 & 1.08 - 1.59 & 177 - 230 & 0.24 - 1 \\
WD\,0137$-$349     & 0.0829738 & 62 - 259  & 1.07 - 1.79 & 48 - 108 & 0.19 - 1 \\
NLTT\,5306         & 0.0827715 & 101 - 582 & 1.09 - 1.84 & 73 - 178 & 0.08 - 1 \\
SDSS\,J1557$+$0916 & 0.0951833 & 104 - 627 & 1.06 - 1.85 & 75 - 181 & 0.06 - 1 \\
    EPIC\,212235321     & 0.0482488 & 249 - 632 & 1.06 - 1.73 & 131 - 181 & >0.03 - 0.29*\\
\hline
\multicolumn{5}{c}{With minimum total age restriction}\\
\hline
SDSS\,J1411$+$2009 & 0.0884218 & 334 - 937 & 1.08 - 1.56 & 153 - 276 & 0.09 - 0.54 \\
SDSS\,J1205$-$0242 & 0.0509304 & 67 - 200  & 1.07 - 1.58 & 54 - 92   & 0.18 - 0.99 \\
WD\,1032$+$011     & 0.0983930 & 139 - 605 & 1.08 - 1.32 & 79 - 178  & 0.06 - 0.41 \\ 
ZTF\,J0038$+$2030  & 0.4323880 & 591 - 925 & 1.08 - 1.14 & 177 - 230 & 0.24 - 0.43 \\ 
WD\,0137$-$349     & 0.0829738 & 62 - 259  & 1.07 - 1.79 & 48 - 108 & 0.19 - 1 \\
NLTT\,5306         & 0.0827715 & 131 - 582 & 1.09 - 1.34 & 78 - 178 & 0.08 - 0.49\\
SDSS\,J1557$+$0916 & 0.0951833 & 104 - 627 & 1.06 - 1.85 & 75 - 181 & 0.06 - 1 \\
EPIC\,212235321     & 0.0482488 & 249 - 632 & 1.06 - 1.73 & 131 - 181 & >0.03 - 0.29*\\
\hline   
\hline
\end{tabular} 
\noindent
\\
*The given range was calculated using the upper limit for the BD mass from the literature, and increases for smaller masses.
\\
\label{tab:res}
\end{table*}

%%%%%%%%%%%%%%%%%%%%%%%%%%%%%%%%%%%%%%%%%%%%%%%%%%
%%%%%%%%%%%%%%%%% RECONSTRUCTION %%%%%%%%%%%%%%%%%
%%%%%%%%%%%%%%%%%%%%%%%%%%%%%%%%%%%%%%%%%%%%%%%%%%

\section{Reconstruction algorithm}

For the reconstruction of the CE phase we used the same method we applied in our previous articles \citep{zorotovicetal10-1,zorotovicetal14-2,parsons2015,Hernandez21,hernandezetal22}, i.e. based on a grid of stellar evolution tracks generated with the Single Star Evolution (SSE) code from \citet{hurley00} for solar metallicity and Roche-geometry \citep[for a detailed explanation of the algorithm we refer the reader to section 4.1 in][]{Hernandez21}. Based on the WD's cooling ages from the literature we calculated the period the systems had immediately after the CE was ejected ($P_{\mathrm{CE}}$), assuming that only gravitational radiation had reduced the orbital angular momentum during the post-CE phase \citep{schreiber+gaensicke03}. The binding and orbital energy were calculated as in the Binary Star Evolution (BSE) code from \citet{hurleyetal2002}, i.e.
\begin{equation}
\label{eq:Ebin}
\Ebin = - \frac{G M_\mathrm{1} M_\mathrm{1,e}}{\lambda R_\mathrm{1}},
\end{equation}
were $M_\mathrm{1}$, $M_\mathrm{1,e}$ and $R_\mathrm{1}$ are the WD's progenitor mass, envelope mass and radius at the onset of the CE phase and $\lambda$ is a structural parameter that depends on its mass distribution and radius, and
\begin{equation}
\label{eq:Deor}
\Delta\Eorb = E_\mathrm{orb,i} - E_\mathrm{orb,f}= \frac{1}{2} G M_\mathrm{1,c} M_\mathrm{2}\left(\frac{1}{a_\mathrm{i}}  - \frac{1}{a_\mathrm{f}}\right),
\end{equation}
where $M_\mathrm{1,c}$ is the core mass of the primary (i.e. the current WD mass), $M_\mathrm{2}$ the mass of the companion and $a_\mathrm{i}$ and $a_\mathrm{f}$ correspond to the orbital separation immediately before and after the CE phase, respectively. The structural parameter $\lambda$ was calculated with the equations developed by Onno Pols, included in the last version of the BSE code and published in \citet{izzardthesis,claeysetal14}. These equations also allowed us to include a fraction of the hydrogen recombination energy (called  ionization energy in \citealt{claeysetal14}) which reduces the binding energy of the envelope by increasing the value of $\lambda$. However, in our reconstruction we assumed that this energy does not contribute to the process (i.e. the fraction was set to zero) in order to test whether the observed systems could be reconstructed without extra energy sources. The maximum total age of the systems for the reconstruction was set to the age of the Galactic disc \citep[i.e. 10\,Gyr,][]{Oswalt96}.

%%%%%%%%%%%%%%%%%%%%%%%%%%%%%%%%%%%%%%%%%%%%%%%%%%
%%%%%%%%%%%%%%%%%     RESULTS    %%%%%%%%%%%%%%%%%
%%%%%%%%%%%%%%%%%%%%%%%%%%%%%%%%%%%%%%%%%%%%%%%%%%

\section{Results}
\label{sec:res}

We were able to reconstruct the evolution of all the detached WD+BD binaries in our sample without the need of recombination energy (or any other extra energy sources) for the ejection process, in the same way as we did for the observational sample of detached WD+dM PCEBs \citep{zorotovicetal10-1} and for most of the WD+AFGK PCEBs \citep{parsons2015,Hernandez21,hernandezetal22}.

In Table\,\ref{tab:res} we present the derived values for the period immediately after the CE phase and ranges derived from our reconstruction algorithm for the initial orbital period, initial mass of the progenitor of the WD, radius of the WD's progenitor at the onset of the CE phase and CE efficiency $\alpha_{\mathrm{CE}}$. We distinguish two sets of results, the first one allowing the total age of the systems to take any value between the cooling age of the WD and the age of the Galactic disc, and the second and more restrictive one including the minimum age constraints mentioned in Sec.\,\ref{sec:sample}. In what follows we describe the results we obtained separating the eclipsing and the non-eclipsing systems. 

\begin{figure*}
 \centering
 \includegraphics[width=0.99\textwidth,angle=0]{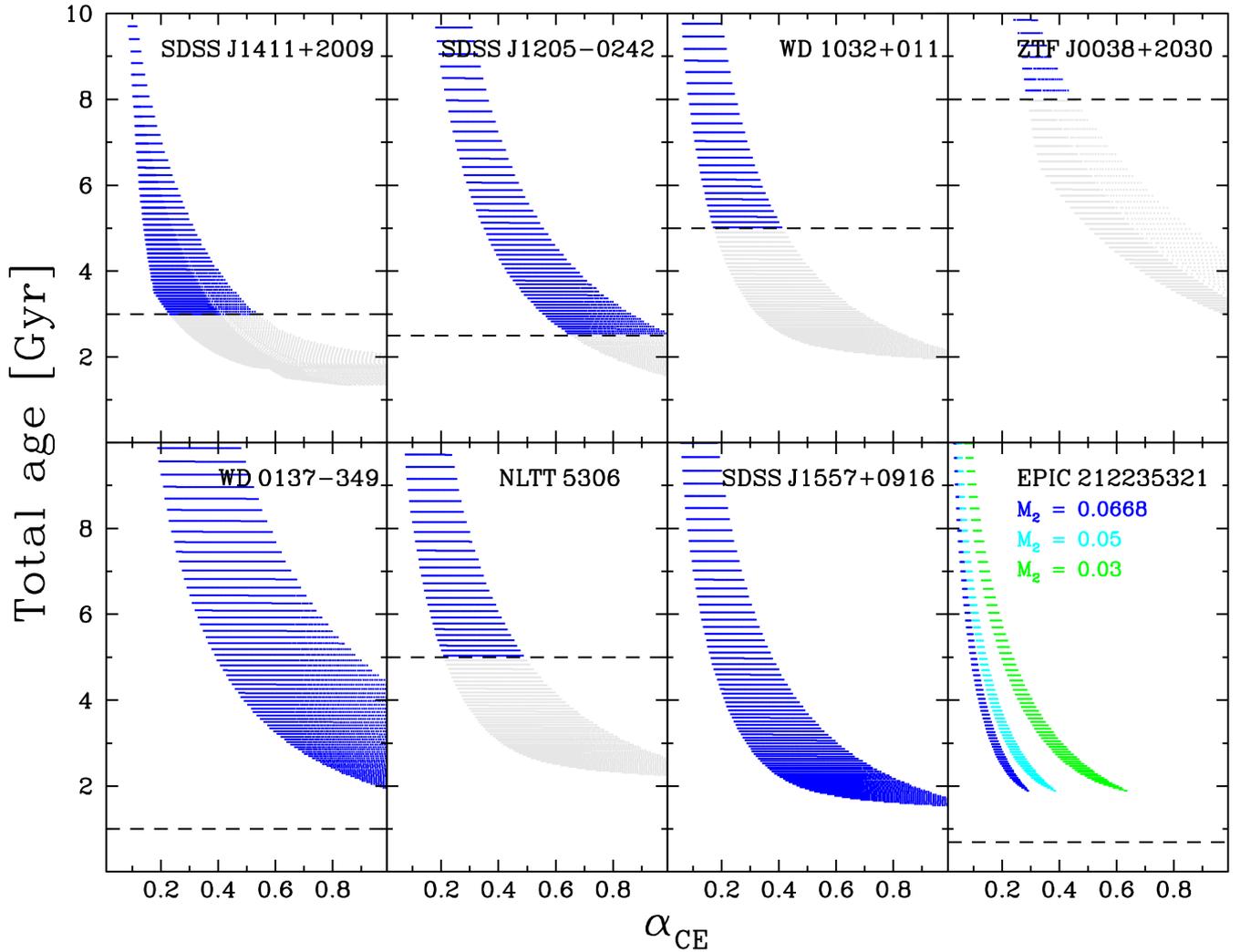}
 \caption{Possible values for the total age reconstructed for the eight WD+BD PCEBs in our sample depending on the CE efficiency without minimum age restriction (gray) and with minimum age from the literature (blue) given by the dashed line in each panel. For EPIC\,21235321 we show the results with three different BD masses given that only an upper limit for the BD mass is given in the literature. While the minimum values of $\alpha_{\mathrm{CE}}$ come from the maximum age (assuming that all the systems belong to the Galactic disc), the maximum value of $\alpha_{\mathrm{CE}}$ for each systems is reduced to less than 1 only after considering the estimated minimum ages, which reduces the range of $\alpha_{\mathrm{CE}}$ for which all the systems can be reconstructed to $0.24 - 0.41$.} 
 \label{fig:1}
 \end{figure*}
 
\subsection{Eclipsing systems}
\label{res:ecl}
 
In Fig.\,\ref{fig:1} we show the results obtained for the total age of the systems (i.e. time until they entered the CE phase plus the cooling age of the WD) as a function of the derived CE efficiency. Gray points correspond to all solutions without a lower limit for the age of the systems (besides a total age larger than cooling age), while blue points correspond to solutions in which the total age is larger that the minimum age from the literature (given by the dashed line in each panel). For the four eclipsing systems (upper panels) we obtained a very wide range of possible values for $\alpha_{\mathrm{CE}}$. The maximum age of 10\,Gyr assumed for all the systems set minimum values of $\alpha_{\mathrm{CE}}=$ 0.09, 0.18, 0.06 and 0.24 for SDSS\,J1411$+$2009, SDSS\,J1205$-$0242, WD\,1032$+$011 and ZTF\,J0038$+$2030, respectively. Without minimum age restriction $\alpha_{\mathrm{CE}}$ goes up to 1 for the four systems. However, the minimum total ages estimated from the literature impose a maximum value for $\alpha_{\mathrm{CE}}$ of 0.54, 0.99, 0.41 and 0.43, respectively. This means that the four systems can be reconstructed with the same value of $\alpha_{\mathrm{CE}} = 0.24 - 0.41$, which is in perfect agreement with the low value derived from observations of detached PCEBs consisting of a WD and a low-mass main-sequence companion \citep{zorotovicetal10-1}, with no evidence for extra energy sources being involved.

Based on the WD masses, the progenitors of the WDs in SDSS\,J1411$+$2009 and ZTF\,J0038$+$2030 most likely filled their Roche lobes when they were on the asymptotic giant branch (AGB), while the WD masses in SDSS\,J1205$-$0242 and WD\,1032$+$011 are more consistent with helium-core WDs, which descend from progenitors that filled the Roche lobe during the first giant branch (FGB). 

\subsection{Non-eclipsing systems}
\label{res:nonecl}

The evolutionary history for all the non-eclisping systems in our sample can also be understood assuming a small CE efficiency and without the inclusion of any additional sources of energy such as recombination energy. Based on the WD masses the WDs in the four non-eclisping systems most likely filled their Roche lobes when they were on the FGB, which typically results in a helium-core WD. However, given the WD mass and the range of progenitor masses derived for EPIC\,212235321, it is very likely that the WD in this system experienced a previous phase of core-helium burning as a hot subdwarf B (sdB) star \citep[see e.g.,][for a review on hot subdwarf stars]{heber2016}. This is not surprising, given the large population of known eclipsing sdB stars with close dM or BD companions (known as HW\,Vir systems) that are most likely PCEBs \citep[see e.g.,][and references therein]{Schaffenroth21}. If the WD in EPIC\,212235321 was indeed an sdB star before becoming a WD,
then its core will be composed of carbon and oxygen. 

As can be seen in Fig.\,\ref{fig:1}, bottom panels, for three of the four systems we found a rather large range of possible efficiencies while the shortest orbital period PCEB in our sample, EPIC\,212235321, requires a small efficiency. However, this strong constraint is largely related to assuming the observationally determined upper limit for the BD mass in our reconstruction algorithm. If the real BD mass is smaller, the upper limit on $\alpha_{\mathrm{CE}}$ increases. For example, for a BD mass of $0.05$\Msun\, we obtained $\alpha_{\mathrm{CE}}=0.04-0.39$, while for a smaller BD mass of $0.03$\Msun\, the allowed range of efficiencies is $0.07-0.63$ (cyan and green results in Fig.\,\ref{fig:1}, respectively).

The observationally derived minimum age estimates might provide additional hints for the evolutionary history of each system. WD\,0137$-$3497 needs to be older than the minimum estimated age of $\sim1$\,Gyr derived by \citet{burleighetal06} for the BD by at least a factor of two to reach $\alpha_{\mathrm{CE}}\leq1$. This implies that the BD was most likely heated by radiation from the young post-CE WD. For NLTT\,5306, the age restriction of 5-10\,Gyr implies $\alpha_{\mathrm{CE}}=0.08 - 0.49$. In the case of SDSS\,J1557$+$0916, no age estimate is provided in the literature but we note that it must be at least $\sim1.5$\,Gyr old for $\alpha_{\mathrm{CE}}\leq1$ and older than $\sim2.2$\,Gyr for having $\alpha_{\mathrm{CE}}\leq0.41$, consistent with the range of $\alpha_{\mathrm{CE}}$ derived for the eclipsing systems. Finally, for EPIC\,212235321 the minimum age restriction of only $0.7$\,Gyr does not provide further constraints to the range of $\alpha_{\mathrm{CE}}$ allowed for its reconstruction. As we already mentioned, the upper limit on $\alpha_{\mathrm{CE}}$ for this system strongly depends on the assumed BD mass, and increases if we assume a mass below the upper limit from the literature. 

\vspace{0.5cm}
\noindent In summary, if we interpret the estimated ages of the BD companions as lower limits on the age of the system, the formation of the eight PCEBs with detached BD companions that we studied can be reproduced by assuming a small CE efficiency ($\alpha_{\mathrm{CE}} = 0.24 - 0.41$) nearly identical to the one that is required to explain the large sample of PCEBs with low mass stellar companions \citep{zorotovicetal10-1} and the majority of PCEBs with earlier spectral type secondary stars \citep{Hernandez21,hernandezetal22}.

\subsection{Comparison with previous studies}

For some of the eight systems in our sample as well as for one alleged sdB+BD system that we did not reconstruct here, constraints on CE evolution have been presented in previous works. In what follows we relate our results to these earlier attempts. 

\subsubsection{SDSS\,J1205$-$0242}

\citet{parsons17} found the WD in this system to be more likely composed of helium, based on the mass radius relation, which is consistent with its progenitor being on the FGB at the onset of the CE phase, as we derived here. They also did a crude estimation of the initial binary properties based on the energy equation for the CE phase deriving an initial orbital period in the range of $60-200$\,d, in perfect agreement with our reconstruction. The WD mass derived by \citet{rappaport17} is sightly larger ($0.415\pm0.028$\,\Msun) but still consistent with a helium WD. They also analysed possible evolutionary scenarios, favoring a recent emergence of the CE phase, an initial period of $\simeq155^{+60}_{-40}$\,d and an initial primary mass of $\simeq1.18\pm0.26$\,\Msun, which is also consistent with our estimation, especially when the minimum age derived for the system is considered. 

Given the short orbital period of the system, close to the period needed to start mass transfer from the companion to the WD, \citet{rappaport17} also evaluated a possible alternative scenario in which the companion was initially more massive and transferred mass to the WD as a cataclysmic variable, which would imply that the system emerged from the CE phase long ago. If this was the case, it would be impossible to reconstruct the CE phase since the initial mass of the companion and the time since the envelope was ejected are crucial parameters for the reconstruction. However, this scenario was considered highly unlikely by the authors, since the WD would have had to experience a large nova eruption that caused the system to detach and the WD to remain hot and thermally bloated (i.e. in a hibernation state,  \citealt{shara86}), but they found no signatures of mass transfer in the spectrum and there is no evidence for a recent nova eruption at the location of the binary \citep{schafter17}. Additional evidence against a previous nova outburst comes from the lack of helium-core WDs in cataclysmic variables, which can be explained if mass transferring systems with low-mass WDs merge due to the consequential angular momentum loss produced by nova eruptions \citep{schreiber16,nelemans16}. Had the system had a more massive companion that underwent mass transfer to the WD, it would likely have merged resulting in a single helium-core WD \citep[e.g.,][]{Zorotovicetal17}. The alternative scenario according to which a cataclysmic variable can detach if a strong magnetic field is generated through the crystallization and rotation driven dynamo \citep{schreiberetal21-1} can be excluded as well, because this scenario requires a crystallizing carbon/oxygen WD.  

\subsubsection{WD\,1032$+$011}

For this system, \citet{Casewell2020} favor a carbon/oxygen core WD with a thin envelope for the primary star, based on the derived mass and radius of the WD, although the parameters are still consistent with both, helium and carbon/oxygen WD models. The authors did not discuss the possible evolutionary history. Our reconstruction, however, only predicts possible progenitors of the WD that filled the Roche lobe during the FGB. This is a direct consequence of assuming a WD mass in the range of 0.4-0.5\,\Msun\, (i.e. within the uncertainty given by \citealt{Casewell2020}). The SSE code from \citet{hurley00} used to generate our progenitors grid only produces core masses larger than $\sim$0.51\,\Msun\, at the base of the AGB for the range of initial masses we are using (stars that can evolve of the main-sequence within the age of the Galactic disc) and solar metallicity, which implies that we can only find progenitors on the AGB for WD masses above this limit. In order to check how the core mass might be affected by metallicity we generated a new grid assuming a moderately low metallicity (z = 0.01), as was suggested by \citet[][]{Casewell2020} for this system ([Fe/H]$\sim$\,-0.3), and found that the minimum core mass at the base of the AGB slightly increases ($\sim$0.515\,\Msun). Therefore, the reconstruction with lower metallicity also predicts only progenitors on the FGB, and we found the same range of possible values for the CE efficiency. Solutions consistent with a post-AGB carbon/oxygen WD can only be obtained if the real WD mass is at least $\sim$0.06\,\Msun\, larger than the value derived by \citet{Casewell2020}. 

Another possibility, which is consistent with the low mass measured and the carbon-oxygen composition, is that the primary filled its Roche lobe close to the tip of the FGB and its core emerged from the CE phase with a mass that is enough to ignite helium, becoming an sdB star, i.e. a helium-core burning star with a very thin hydrogen envelope. The sdB phase is expected to last for a short time ($\sim$10\,Myr) until helium is exhausted in the core and the star becomes a carbon/oxygen WD. Indeed, based on \citet[][its Figure\,1]{han2002}, the measured WD mass is consistent with this scenario for the small progenitor mass we derived, especially when the minimum age restriction is taken into account. Therefore, we conclude that the WD in this binary is very likely a post-sdB star. 

\subsubsection{ZTF\,J0038$+$2030}

For ZTF\,J0038$+$2030, \citet{vanroestel21} did not reconstruct the CE phase but suggested a WD progenitor mass of $\sim$\,1-2\,\Msun\, based on the initial-to-final mass relation. Taking into account the minimum age estimated for the BD ($\gappr$8\,Gyr), which implies a long main-sequence lifetime for the primary, our reconstruction algorithm is favoring masses closer to 1\,\Msun. 

\citet{vanroestel21} also mentioned that the WD is most likely composed of carbon and oxygen based only on the estimated mass. They discussed two possible scenarios for its evolution, either that the CE phase occurred when the progenitor of the WD was on the AGB or close to the tip of the FGB. The later would imply a previous sdB phase where helium is burned into carbon and oxygen as discussed for WD\,1032$+$011. Restricting the WD mass to lie within the error listed by \citet[][i.e. 0.50$\pm$0.02\,\Msun]{vanroestel21} and assuming solar metallicity we only found possible progenitors on the AGB. However, the maximum core mass at the tip of the FGB in our grid is very close to the minimum WD mass of 0.48\,\Msun\, from the literature, which implies that our algorithm would find progenitors on the FGB by assuming a slightly larger uncertainty of the WD mass (i.e. 0.03\,\Msun\, instead of 0.02\,\Msun). 
Assuming a lower metallicity of the WD progenitor would imply a slightly larger (but still close to 0.48\,\Msun) core mass at the tip of the FGB according to the SSE code. As the system is expected to be old but still in the solar neighbourhood, we tested for z = 0.01 instead of our default (z = 0.02) but still did not get progenitors on the FGB phase within the error estimated by \citet{vanroestel21}. The range of $\alpha_{\mathrm{CE}}$ is only slightly increased for the reduced metallicity (we get 0.21 to 1 for any age, and 0.21 to 0.44 when the minimum age is considered), which does not affect our main conclusion that all the WD+BD systems can be reproduced with the same small CE efficiency and without extra energy sources. 

\subsubsection{WD\,0137$-$349}

WD\,0137$-$349 was discovered as the first WD+BD PCEB by \citet{maxtedetal06}. In a quick follow up paper, \citet{burleighetal06} showed that if the age of the system is the cooling age of the BD ($1$\,Gyr), then the progenitor of the WD must have been relatively massive ($\simeq2\Msun$) and a large CE efficiency $(\alpha_{\mathrm{CE}}\geq2)$ is needed to understand the formation of the system through CE evolution. This result in in agreement with our reconstruction. However, as already mentioned by \citet{burleighetal06}, the cooling age of the BD represents just a lower limit on the total age of the binary star system as it could have been heated significantly following CE evolution (especially at early times when the WD was very hot). Interpreting the cooling age of the BD as a lower limit of the total age of the binary, we find solutions with small values of the CE efficiency. Assuming an age of just $\sim2$\,Gyr allows us to find solutions with $\alpha_{\mathrm{CE}}\leq1$, however the system must be at least $\sim5$\,Gyr old in order to be consistent with the small value of $\alpha_{\mathrm{CE}}$ ($\leq0.41$) derived for the eclipsing WD+BDs in the sample. 

\citet{davis2012} analysed PCEBs with WD primaries using a statistical approach deriving a weighted mean of $\alpha_{\mathrm{CE}}$ over all possible solutions. The only PCEB with a sub-stellar companion in their sample was WD\,0137$-$349\footnote{named WD\,0137$-$3457 in their tables}. Based on their reconstruction, which did not take into account any age constraints, \citet{davis2012} claimed that PCEBs with BD companions require a larger value of the CE efficiency and that this value is exceeding one. In contrast, we here reconstructed the evolutionary history of WD\,0137-349 using values of $\alpha_{\mathrm{CE}}=0.19-1$. This shows that there are no indications for a larger efficiency for PCEBs with BD companions. We speculate that the difference between our result and those presented by \citet{davis2012} lies partly in the very sparse grid of structural parameters ($\lambda$) from \citet{dewitauris2000} that was used by \citet{davis2012}.

\subsubsection{HD 149382} 

We did not reconstruct the evolutionary history of the alleged PCEB with a BD companion HD\,149382 because the primary star is an sdB star and not a WD. The evolutionary history of sdB stars with close dM or BD companions will be discussed in a forthcoming work. However, we need to briefly mention this system here as in a previous work strong claims have been made based on its alleged existence. 

\citet{demarco11} argue that to reconstruct PCEBs with BD companions requires an increased CE efficiency based on just this one system using the parameters derived by \citet{geier09}. However, in subsequent studies of HD\,149382, it has been shown that the existence of a close sub-stellar companion can be excluded for orbital periods below 28 days and $Msini\,\gappr\,1\,\Mjup$ \citep{norris11,jacobs11}. Therefore, the claims for an increased CE efficiency for PCEBs with sub-stellar companions based on HD\,149382 should be discarded.

%%%%%%%%%%%%%%%%%%%%%%%%%%%%%%%%%%%%%%%%%%%%%%%%%%
%%%%%%%%%%%%%%%%%     SUMMARY    %%%%%%%%%%%%%%%%%
%%%%%%%%%%%%%%%%%%%%%%%%%%%%%%%%%%%%%%%%%%%%%%%%%%

\section{Concluding discussion}

We have compiled a list of observationally well characterized white dwarf plus brown dwarf (WD+BD) post common envelope binaries (PCEBs) and reconstructed their formation history using the energy formalism for common-envelope (CE) evolution which provides constraints on the CE efficiency parameter $\alpha_{\mathrm{CE}}$. We find reasonable solutions for all systems in our sample within the energetically allowed rage of $\alpha_{\mathrm{CE}}\leq1$ without considering contributions from recombination energy. If we take into account lower limits on the total age of the systems observationally derived from the brown dwarf companions or kinematics of the binary stars, we find that the formation of all WD+BD systems in our sample can be reproduced assuming $0.24\leq\alpha_{\mathrm{CE}}\leq0.41$. This finding differs drastically from previous claims of an increased efficiency for PCEBs with sub-stellar companions \citep{demarco11,davis2012} but is in very good agreement with previous results obtained for PCEBs consisting of white dwarfs with main sequence star companions \citep{zorotovicetal10-1,Hernandez21,hernandezetal22}. 

Despite the fact that the formation of the vast majority of PCEBs can therefore be understood assuming a small CE efficiency and without taking into account additional energy sources, for three systems with observationally well constrained parameters the situation is different. IK\,Peg and KOI-3278 are potentially PCEBs with orbital periods significantly longer than that of typical PCEBs and can only be understood if energy sources apart from orbital energy help to expel the envelope. However, as shown by \citet{zorotovicetal14-2}, only two percent of the recombination energy stored in the envelope could provide the required extra-energy. The third example of a potential PCEB that cannot be understood if only orbital energy is considered, is given by the giant planet orbiting the white dwarf WD\,1856+543 \citep{vanderburgetal20-1}. In this case it appears that a higher fraction of recombination energy ($\gappr10$\,per cent), somewhat depending on the relatively uncertain white dwarf mass measurement, is required to understand its formation through CE evolution \citep{lagosetal21-1}. As an alternative, \citet{lagosetal21-1} and \citet{chamandyetal21-1} suggested that the envelope could have been expelled by several planets of which just the one we observe today survived. 

To summarize, using the CE energy equation, assuming a small efficiency, and ignoring possible contributions from recombination energy, allows to reproduce the dominant population of PCEBs. In rare individual cases, uncertain additional energy sources are required.

%%%%%%%%%%%%%%%%%%%%%%%%%%%%%%%%%%%%%%%%%%%%%%%%%%
%%%%%%%%%%%%%%%%%     ACK    %%%%%%%%%%%%%%%%%
%%%%%%%%%%%%%%%%%%%%%%%%%%%%%%%%%%%%%%%%%%%%%%%%%%

\section*{Acknowledgements}
MZ and MRS acknowledge support from Fondecyt (grant 1221059). MRS also acknowledges ANID, -- Millennium Science Initiative Program -- NCN19\_171. We thank the referee, Sarah Casewell, for her constructive suggestions, which helped us to improve this article.

\section*{Data Availability}
The data used in this article can be obtained upon request to the corresponding author and after agreeing to the terms of use.

\bibliographystyle{mnras}
\bibliography{WDBD} % if your bibtex file is called example.bib

% Don't change these lines
\bsp	% typesetting comment
\label{lastpage}
\end{document}